\documentclass[fleqn,usenatbib]{mnras}
\usepackage{newtxtext}
\usepackage[T1]{fontenc}
\usepackage{graphicx}
\volume{{\rm in press}}

\newcommand{\teff}{${T}_{\mathrm{eff}}$}
\newcommand{\logg}{$\log{g}$}
\newcommand{\msun}{$M_{\odot}$}
\newcommand{\kep}{{\em Kepler}}

\newcommand{\ktwo}{{\em K2}}

\newcommand{\muhz}{$\mu$Hz}

\title[Caveats in using white dwarfs as flux standards]{When flux standards go wild: white dwarfs in the age of {\em Kepler}}

\author[Hermes et al.]{J.~J.~Hermes,$^{1}$\thanks{jjhermes@unc.edu}\thanks{Hubble Fellow}
B.~T.~G\"{a}nsicke,$^{2}$
Nicola~Pietro~Gentile~Fusillo,$^{2}$
R.~Raddi,$^{2}$
\newauthor
M.~A.~Hollands,$^{2}$
E.~Dennihy,$^{1}$
J.~T.~Fuchs,$^{1}$
and S.~Redfield$^{3}$
\\
$^{1}$Department of Physics and Astronomy, University of North Carolina, Chapel Hill, NC\,-\,27599-3255, USA\\
$^{2}$Department of Physics, University of Warwick, Coventry\,-\,CV4~7AL, UK\\
$^{3}$Wesleyan University Astronomy Department, Van Vleck Observatory, 96 Foss Hill Drive, Middletown, CT\,-\,06459, USA\\
}

\begin{document}

\maketitle

\label{firstpage}

\begin{abstract}

White dwarf stars have been used as flux standards for decades, thanks to their staid simplicity. We have empirically tested their photometric stability by analyzing the light curves of 398 high-probability candidates and spectroscopically confirmed white dwarfs observed during the original {\em Kepler} mission and later with {\em K2} Campaigns 0$-$8. We find that the vast majority ($>$97\,per\,cent) of non-pulsating and apparently isolated white dwarfs are stable to better than 1\,per\,cent in the {\em Kepler} bandpass on 1-hr to 10-d timescales, confirming that these stellar remnants are useful flux standards. From the cases that do exhibit significant variability, we caution that binarity, magnetism, and pulsations are three important attributes to rule out when establishing white dwarfs as flux standards, especially those hotter than $30\,000$\,K.

\end{abstract}


% ++++++++++++++++++++++++++++++++++++++++++++++++++++++++++++++++++++ %
\section{Introduction}
\label{sec:intro}

Accurate, reliable flux standards are essential for the calibration of absolute photometry and spectroscopy. Many of the most delicate astrophysical observations are limited by systematic uncertainties in basic flux calibration, most notably next-generation surveys to more accurately measure dark energy using supernovae (see \citealt{2015MPLA...3030030S}, and references therein).

Typically, atmospheric variability and instrumental artifacts dominate calibration errors \citep{2006ApJ...646.1436S}. However, inherent stellar variability can propagate into the uncertainties if unsuitable standards are chosen.

Hot, hydrogen-atmosphere (DA) white dwarfs ($18\,000-80\,000$\,K) have been used as standards for decades: they are close, minimizing interstellar reddening, and have relatively simple, purely radiative atmospheres that can be described completely by their effective temperature and surface gravity \citep{2016ApJ...822...67N}. The {\em Hubble Space Telescope} CALSPEC standard star network is anchored to three hot DAs: G191-B2B, GD\,153, and GD\,71 \citep{2007ASPC..364..315B}. An identical or similar sample of white dwarfs (and additional cooler stars) is expected to calibrate the next major space observatory, the {\em James Webb Space Telescope} \citep{2011AJ....141..173B}.

We know empirically that not all white dwarfs are suitable flux standards. Cooler DA white dwarfs were originally used for flux calibration, but that changed with the discovery that those with convective atmospheres showed photometric variability up to several per\,cent on the timescale of minutes \citep{1968ApJ...153..151L}; these are oscillations in the variable DA (ZZ Ceti) stars, which pulsate when they cool to between roughly $12\,500-10\,500$\,K \citep{2008ARA&A..46..157W}. Additionally, strongly magnetic white dwarfs with convective atmospheres have shown large-amplitude, rotational variability (e.g., \citealt{2013ApJ...773...47B}).

However, we so far have few empirical constraints on the stability of hot white dwarfs. That has changed with the revolution in long-term monitoring enabled by the {\em Kepler} space telescope, which was launched to discover Earth-like planets around Sun-like stars. {\em Kepler} data is precise enough to deliver tens of parts-per-million photometry on thousands of bright stars \citep{2013Natur.500..427B}, and has been used to detect low-level variability in a handful of the 14 non-pulsating white dwarfs observed in the original {\em Kepler} mission \citep{2015MNRAS.447.1749M}.

After the failure of the second reaction wheel, the {\em Kepler} spacecraft has been repurposed as {\em K2}, surveying new fields along the ecliptic plane roughly every three months \citep{2014PASP..126..398H}. This has dramatically increased the number of white dwarfs available for extended monitoring from space; hundreds of known and candidate white dwarfs have been observed to look for eclipses \citep{2016MNRAS.458..845H} and transits \citep{2015Natur.526..546V}, as well as to perform asteroseismology \citep{2014ApJ...789...85H}.

% ++++++++++++++++++++++++++++++++++++++++++++++++++++++++++++++++++++ %
\begin{figure*}
\centering{\includegraphics[width=0.88\textwidth]{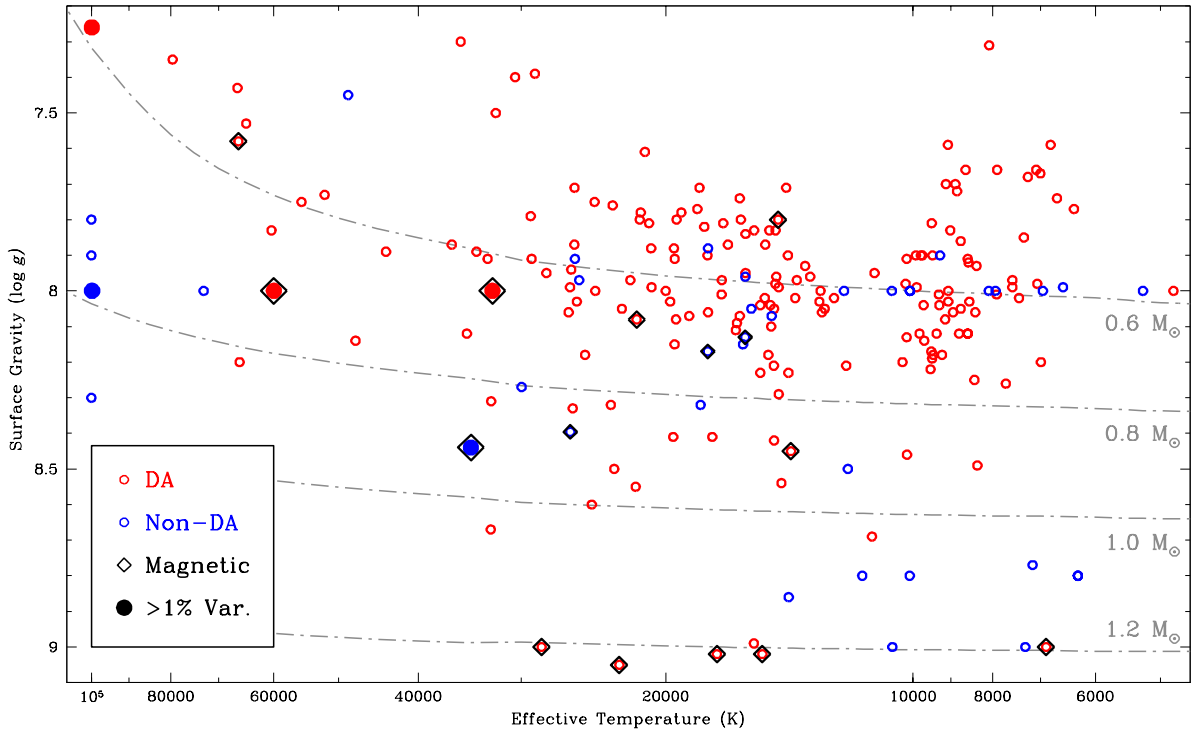}}
\caption{The ${T}_{\mathrm{eff}}-\log{g}$ plane for 252 spectroscopically confirmed white dwarfs observed through \ktwo\ Campaign 8 brighter than $K_{\rm p}$$<$$19.0$\,mag, with DA cooling tracks \citep{2001PASP..113..409F} plotted to guide the eye. The small circles, coloured by spectral class, identify the 245 white dwarfs ($>$$97$\,per\,cent) suitable as flux standards, with maximal variability amplitudes $<$$10$\,ppt ($<$$1$\,per\,cent) in the \kep\ bandpass (roughly SDSS-$r$). We have excluded here all pulsating white dwarfs and those with detected line-of-sight companions. We highlight in filled circles the large-amplitude variables that would be poor flux standards (see Figure~\ref{fig:badflux}). Three are likely magnetic white dwarfs; the other two are hotter than 90,000 K, and we are likely seeing reflection from a close companion. \label{fig:loggteff}}
\end{figure*}
% ++++++++++++++++++++++++++++++++++++++++++++++++++++++++++++++++++++ %

We report here an analysis of the first 252 spectroscopically confirmed, non-pulsating, and apparently single white dwarfs observed in the original {\em Kepler} mission and subsequently with {\em K2} through Campaign 8, as well as 146 high-probability white dwarf candidates without spectroscopy. Our observations and analysis are outlined in Sections \ref{sec:obs} and \ref{sec:variability}, and we detail caveats in the use of white dwarfs as flux standards from what we have learned so far from {\em Kepler} in Section~\ref{sec:caveats}. We conclude in Section~\ref{sec:discussion}.

\section{Observations and Analysis}
\label{sec:obs}

\subsection{Target Selection}
\label{sec:obs:selection}

White dwarfs in the original {\em Kepler} mission were targeted by a search for compact objects, and characterized spectroscopically by \citet{2010MNRAS.409.1470O,2011MNRAS.414.2860O}. We requested additional {\em K2} observations of known and candidate white dwarfs through various Guest Observer programs, accepted in each pointing of Campaigns $0-8$. The majority of white dwarfs with spectroscopic information were discovered from the Sloan Digital Sky Survey (SDSS, \citealt{2013ApJS..204....5K}). Additionally, we proposed many candidates with colours and reduced-proper-motions selected from SDSS photometry, with high probabilities of being a white dwarf ($P_{\rm WD}>0.7$), as defined by \citet{2015MNRAS.448.2260G}.

We removed 34 white dwarfs from our sample with significant short-period variability detected from pulsations, all of which were proposed by us and observed by {\em Kepler} in short cadence every 58.8\,s. Additionally, we cross-matched our sample with the most recent WD+MS catalog of \citet{2016MNRAS.458.3808R} as well as DA+dM pairs from the ESO Supernova Ia Progenitor Survey \citep{2009A&A...505..441K}. This removed 72 white dwarfs with spectroscopic evidence of a line-of-sight, main-sequence companion; many of these systems are post-common-envelope binaries (PCEBs) and show photometric modulation from reflection from a close companion (e.g., \citealt{2010MNRAS.402.2591P}). We also removed two known eclipsing, single-lined PCEBs: EPIC\,201649211 (SDSSJ1152+0248, \citealt{2016MNRAS.458..845H}) and EPIC\,210659779 (NLTT\,11748, \citealt{2010ApJ...716L.146S}).

This left 252 spectroscopically confirmed, non-pulsating, and apparently isolated white dwarfs, after selecting only those brighter than $K_{\rm p} < 19.0$\,mag in the {\em Kepler} bandpass; long-term instrumental systematics dominate the fainter objects. The full distribution of spectral classifications, effective temperatures, and surface gravities is represented in Figure~\ref{fig:loggteff}. Our sample, which includes targets observed in the original {\em Kepler} mission, includes 15 white dwarfs with strong magnetic fields (DAH or DBH), detected from Zeeman splitting. Most white dwarfs are DA, but there are many with helium-dominated (DB and DO), carbon-dominated (DQ), or continuum-dominated (DC) atmospheres, which we classify in Figure~\ref{fig:loggteff} as non-DA.

% +++++++++++++++++++++++++++++ Table 1 +++++++++++++++++++++++++++++ %
\begin{table*}
 \centering
  \caption{White dwarfs observed to be poor flux standards by \kep\ and \ktwo. We mark with a $^{\dagger}$ those with short-cadence data.  \label{tab:badflux}}
  \begin{tabular}{@{}lcccccccrrr@{}}
  \hline
  KIC/EPIC & \ktwo\ & $K_{\rm p}$ & RA  &  Dec  &  Spec. & \teff\ & \logg\     & Period    & Amp.  &  Time of Minimum    \\
         & Field  & (mag)         &  (J2000)  & (J2000) & Class & (K) & (cm s$^{-1}$)   & (hr)      & (per\,cent) &  (BJD$_{\rm TDB}-2456000$)  \\
 \hline
9535405$^{\dagger}$   & K1 & 17.4 & 19 41 31.33 & $+$46 06 10.8 & DAH & $34\,000$ & 8.00 & 6.1375030(13) & 4.404(53) &  1010.80424(31) \\
211719918$^{\dagger}$ & C5 & 15.7 & 08 56 18.95 & $+$16 11 03.8 & DBH & $34\,520$ & 8.44 & 5.706259(12)  & 4.273(24) &  1176.9239(15)  \\
211995459$^{\dagger}$ & C5 & 18.6 & 08 43 30.81 & $+$20 10 49.1 & DAH & $60\,000$ & 8.00 & 53.351(15)    & 5.47(29) &  807.124479(71)  \\
206197016 & C3 & 16.5 & 22 46 53.73 & $-$09 48 34.5 & DA  & $99\,900$ & 7.26 & 19.89770(29)  & 6.391(15) &  1176.749980(46)  \\
228682372 & C5 & 18.6 & 08 39 59.93 & $+$14 28 58.0 & DO  & $99\,800$ & 5.04 & 11.45902(79)  & 2.752(53) &  1176.2302(63)  \\
206473386 & C3 & 18.6 & 22 21 42.49 & $-$05 23 49.8 &   & $\sim$$7750$ &     & 199.54(0.31) & 3.114(73) &  1006.969(31)  \\  %u-g,g-r: 0.49,0.10
210609259$^{\dagger}$ & C4 & 17.7 & 03 44 31.03 & $+$17 05 43.9 &   & $\sim$$8750$ &     & 48.9816(39)  & 3.648(18) &  1096.6434(16)  \\  %u-g,g-r: 0.42,0.02
220306617 & C8 & 18.9 & 01 03 31.68 & $+$02 46 36.0 &   & $\sim$$7750$ &     & 119.14(24)   & 1.79(10)  &  1427.444(46)  \\  %u-g,g-r: 0.44,0.11
220333558 & C8 & 18.7 & 01 01 36.20 & $+$03 21 02.7 &   & $\sim$$8750$ &     & 29.529(14)   & 1.046(57) &  1430.650(11)  \\  %u-g,g-r: 0.40,0.03
\hline
\end{tabular}
\end{table*}
% ==================================================================== %

In addition to the 252 with spectroscopy, several hundred candidate white dwarfs with $K_{\rm p}<19.0$\,mag have been proposed through {\em K2} Campaign 8 without spectroscopy. Some have been proposed from various catalogs of candidate white dwarfs (e.g., \citealt{2011MNRAS.417...93R}, \citealt{2011AJ....142...92B}), but we exclude many here because we do not have sufficient colour and/or proper-motion information to have high confidence they are in fact white dwarfs. However, we expand our sample using targets with SDSS colours consistent with white dwarfs, as well as high reduced proper motions. We inspect only those with probabilities of being white dwarfs exceeding $P_{\rm WD}>0.7$, as defined by \citet{2015MNRAS.448.2260G}, yielding an additional 146 targets for analysis. This brings our total sample to 398 targets.

\subsection{Space-Based Photometry}
\label{sec:obs:K2}

In all cases, we have initially analyzed only the long-cadence data, which are collected by the {\em Kepler} spacecraft every 29.4\,min. In four targets with $>$1\,per\,cent variability (marked with a dagger by the KIC or EPIC identifier in Table~\ref{tab:badflux}) we have analyzed the available short-cadence data collected every 58.8\,s.

Our light curves from the original {\em Kepler} mission were processed by the Kepler Asteroseismic Science Operations Center using Data Release 25 \citep{2014MNRAS.445.2698H}. The {\em K2} data require more care. Using just two reaction wheels for pointing, the spacecraft checks its roll orientation roughly every 6\,hr, and if solar pressure has caused enough of a deviation, {\em Kepler} counteracts its drift by firing its thrusters; this causes significant discontinuities in the photometry. Several pipelines have been developed to process {\em K2} data, but we use here exclusively light curves produced by the {\sc K2sff} routine \citep{2014PASP..126..948V} as well as the Guest Observer office \citep{2016PASP..128g5002V}. Comparing both independently processed light curves for each target, we choose the one that minimizes signal at the thruster-firing timescale, careful to ensure the reduction has the smallest possible aperture to enclose only our white dwarf target. We performed an iterative clip of all points more than 5$\sigma$ discrepant from the median to produce a final light curve.

The majority of light curves have long-term systematics on $10-20$\,d timescales, to varying amplitudes depending on the magnitude of the target. These long-term trends are due to a variety of reasons (see discussion in Section~4 of \citealt{2016ApJ...829...82B}), most commonly from thermal variations on board the spacecraft.

We have computed a Lomb-Scargle periodogram for each light curve, excluding the regions within 0.25\,\muhz\ of all harmonics of the thruster-firing timescale (47.2\,\muhz), as well all signals below 1.157\,\muhz\ (with periods longer than 10\,d). We discuss here those with total amplitudes of variability at a constant period exceeding 1\,per\,cent in the {\em Kepler} bandpass.

\section{Overall White Dwarf Flux Stability}
\label{sec:variability}

Seven of the 252 spectroscopically confirmed white dwarfs observed by {\em Kepler}, spanning the original mission through {\em K2} Campaigns 0$-$8, show peak-to-peak photometric variability exceeding 1\,per\,cent amplitude. We note that several dozen more white dwarfs in our sample show significant variability but to amplitudes below 1\,per\,cent, such that their overall intrinsic photometric stability would still make them decent flux standards.

% ++++++++++++++++++++++++++++++++++++++++++++++++++++++++++++++++++++ %
\begin{figure}
\centering{\includegraphics[width=0.95\columnwidth]{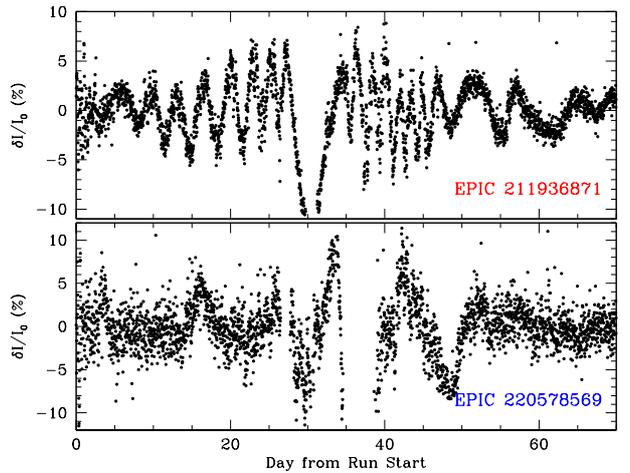}}
\caption{Unsmoothed light curves showing the first 70\,d of two targets with large-scale instrumental artifacts, likely caused by time-varying bias changes, often referred to as rolling bands. We plot EPIC\,211936871 ($K_{\rm p}=18.5$\,mag, Campaign 5) above and EPIC\,220578569 ($K_{\rm p}=18.9$\,mag, Campaign 8) below. Data from both targets were read out from Channel 26 (Module 9.2), which is known to suffer from rolling band pattern noise. Both objects were excluded from our analysis of white dwarf flux stability. \label{fig:rollingbands}}
\end{figure}
% ++++++++++++++++++++++++++++++++++++++++++++++++++++++++++++++++++++ %

% ++++++++++++++++++++++++++++++++++++++++++++++++++++++++++++++++++++ %
\begin{figure*}
\centering{\includegraphics[width=0.95\textwidth]{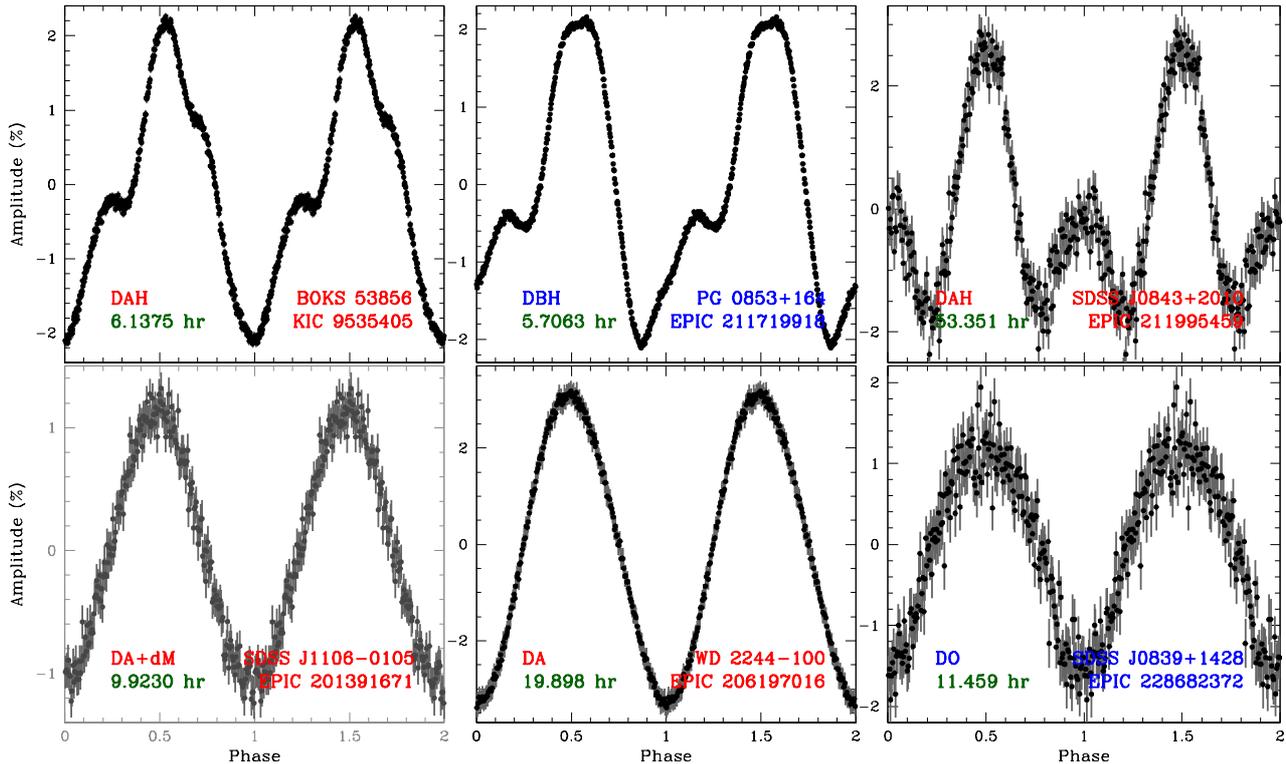}}
\caption{Folded light curves of the five spectroscopically confirmed white dwarfs observed by the {\em Kepler} spacecraft showing $>$$1$\,per\,cent photometric variability. The three white dwarfs at top all have claimed detections of surface magnetic fields, which are likely causing variability at the white dwarf rotation period. The three bottom targets are likely short-period binaries showing reflection from a close companion at the orbital period. The target at the bottom left, EPIC\,201391671, is a known line-of-sight WD+dM system excluded from our sample since the dM is detected spectroscopically \citep{2016MNRAS.458.3808R}, but shown here as an example. The two other targets have white dwarfs with \teff\ $\sim100\,000$\,K and significantly outshine a putative companion. \label{fig:badflux}}
\end{figure*}
% ++++++++++++++++++++++++++++++++++++++++++++++++++++++++++++++++++++ %

However, two of these seven white dwarfs show large-scale variability likely due to instrumental effects rather than intrinsic stellar variability. The light curves of EPIC 211936871 (SDSSJ085025.84$+$191639.5, a $15\,990$\,K DA) and EPIC 220578569 (SDSSJ010901.58$+$083354.7, a $16\,000$\,K DB), shown in Figure~\ref{fig:rollingbands}, feature variability that arises from electronic interference artifacts caused by time-varying crosstalk, often referred to as rolling bands \citep{2014AAS...22412007C}. While observed more than 8 months apart in two separate {\em K2} campaigns, both targets were read out from Channel 26 (Module 9.2), known to suffer from rolling band pattern noise\footnote{Rolling bands manifest as time-varying bias changes, caused by crosstalk between the fine-guidance-sensor CCDs and a high-frequency amplifier oscillation in some of the readout channels of the {\em Kepler} science CCDs. Rather than directly correct these time-variable bias changes, exposures exhibiting rolling bands in the original mission were flagged by the {\em Kepler} science team. However, flagging has been discontinued for {\em K2} \citep{2016PASP..128g5002V}. The three readout ports with the worst rolling band patterns are Channel 26 (Module 9.2), Channel 44 (Module 13.4), and Channel 58 (Module 17.2), although the artifact can affect more than 30 of the 84 science CCDs (\citealt{2010SPIE.7742E..1GK}; G. Barentsen, private communication).}. These high-amplitude trends are also seen in pixels extracted outside the target aperture; therefore, we have omitted these two white dwarfs from further analysis.

This leaves five apparently isolated white dwarfs with coherent stellar variability exceeding a peak-to-peak amplitude of 1\,per\,cent, out of the 250 spectroscopically confirmed targets suitable for inspection. We display their light curves in Figure~\ref{fig:badflux}, folded into 200 phase bins at the dominant period of variability and repeated for clarity. Targets with short-cadence photometry (marked with a $^{\dagger}$ symbol in Table~\ref{tab:badflux}) have been folded into 400 phase bins. Table~\ref{tab:badflux} details information about the five spectroscopically confirmed white dwarfs, on which we comment further in Section~\ref{sec:caveats}.

\begin{figure*}
\centering{\includegraphics[width=0.6333\textwidth]{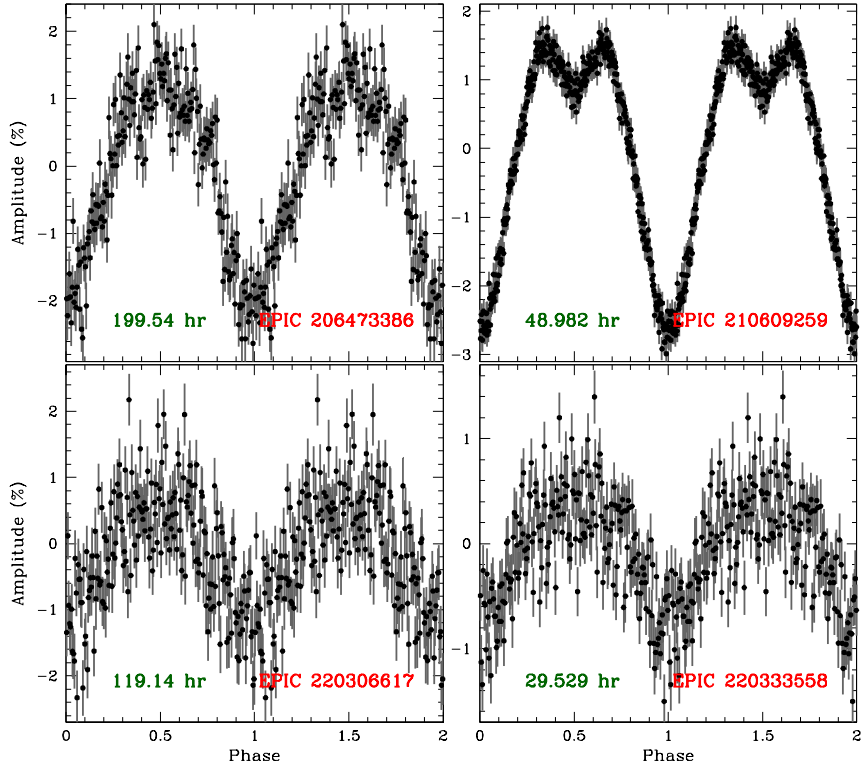}}
\caption{Four white dwarf candidates that are unsuitable flux standards; these targets have high-probability of being white dwarfs from SDSS colours and proper motions \citep{2015MNRAS.448.2260G}. If all are white dwarfs, they likely have convective atmospheres and show modulation at the rotation period. \label{fig:badfluxnospec}}
\end{figure*}

Additionally, we have inspected the light curves of 146 high-probability white dwarfs. Within this subsample, four objects show large-amplitude variability that would make them unsuitable flux standards. All four have photometric colours suggesting they have fully convective atmospheres, with \teff\ $<9000$\,K, and periods of variability exceeding 1\,day. We detail these targets at the end of Table~\ref{tab:badflux}, and show their folded light curves in Figure~\ref{fig:badfluxnospec}.

% ++++++++++++++++++++++++++++++++++++++++++++++++++++++++++++++++++++ %
% ++++++++++++++++++++++++++++++++++++++++++++++++++++++++++++++++++++ %

Overall, we find empirically that just nine of our 396 white dwarf targets (five with spectroscopy and four colour selected) show $>$$1$\,per\,cent amplitude photometric variability. Thus, more than $97$\,per\,cent of our white dwarfs are suitable flux standards.

We note that our analysis is less sensitive to phenomena acting on timescales much shorter than the 30-min cadence of the {\em Kepler} long-cadence photometry. For example, we detect the significant variability caused by transits of the white dwarf EPIC\,201563164 (WD\,1145+017, $K_{\rm p}=17.3$\,mag); this metal-polluted white dwarf is being transited by one or more disintegrating planetesimals \citep{2015Natur.526..546V}. However, the maximum peak of recurrent variability in a periodogram occurs at 4.49\,hr (with 0.76\,per\,cent total amplitude); the deep transits of WD\,1145+017 were smeared out by the 29.4-min cadence of the {\em K2} photometry, and evolved in depth over the campaign. So far, WD\,1145+017 remains the only case of transits we have detected in the nearly 400 single white dwarfs observed through {\em K2} Campaign 8.

\section{Caveats: Binarity, Magnetism, Pulsations}
\label{sec:caveats}

\subsection{Binarity}
\label{sec:binarity}

White dwarfs are not just signposts for the endpoints of stellar evolution, but they also mark the endpoints of binary evolution. Many evolved binaries underwent common-envelope evolution, which brings the orbits closer together to form a PCEB. More than 100 of these WD+MS systems are known, with orbital periods ranging from 1.9\,hr to 4.3\,d \citep{2011A&A...536A..43N}; many show photometric variability at the orbital period \citep{2016MNRAS.461.2747K}.

For this reason, we have removed from our sample all white dwarfs with line-of-sight main-sequence companions, many of which are unresolved within SDSS and could be in close binaries. As described in Section~\ref{sec:obs:selection}, we have excluded all spectroscopically identified WD+dM systems. The analysis of PCEBs in {\em K2} will be discussed in a forthcoming publication.

As an example, we show in the bottom left panel of Figure~\ref{fig:badflux} a known WD+dM system with {\em K2} observations, EPIC\,201391671 (HE\,1103$-$0049). Decomposed fits to the spectroscopy from SDSS show this is a $30\,070\pm190$\,K, $0.41\pm0.02$\,\msun\ white dwarf with a line-of-sight M3 companion \citep{2012MNRAS.419..806R}. The {\em K2} data show a sinusoidal signal (2.13\,per\,cent amplitude) at 9.923\,hr, which arises from a reflection effect off the irradiated face of the M dwarf at the orbital period.

The bottom panel of Figure~\ref{fig:badflux} includes two very hot white dwarfs observed in {\em K2} that have SDSS data but no obvious spectroscopic evidence of a line-of-sight companion. The variability maintains a constant amplitude for $>$$70$\,d with minimal harmonics, suggesting it most likely arises due to irradiation of a companion.

One of the hottest targets in our sample, EPIC\,228682372 (SDSSJ083959.93$+$142858.0), is a DO white dwarf with \teff\ $=99\,800$\,K \citep{2013ApJS..204....5K}. The stable 11.459-hr photometric variability we see from {\em K2} is likely orbital modulation, with the companion outshone by this very hot white dwarf.

Similarly, the hot DA EPIC\,206197016 (WD\,2244$-$100) has \teff\ $=99\,900$\,K and a mass near the canonical mean mass of white dwarfs, $0.59\pm0.03$\,\msun\ \citep{2011ApJ...730..128T}. The sinusoidal photometric variations at 19.898\,hr are most likely caused by reflection off a close companion outshone by this young white dwarf. Infrared photometry in the $YJK$ bands from the VISTA Hemisphere Survey \citep{2013Msngr.154...35M} as well as in band $W1$ from the Wide-Field Infrared Survey Explorer \citep{2010AJ....140.1868W} show an excess of flux from what is expected from a single $100\,000$\,K white dwarf, strongly suggestive of a line-of-sight companion.

To further test this hypothesis, we obtained multi-epoch spectroscopy of EPIC\,206197016 to check for radial-velocity variations. Using the Goodman spectrograph \citep{2004SPIE.5492..331C} on the 4.1-m SOAR telescope, we monitored the velocity of H$\alpha$ over consecutive nights more than 25.6\,hr apart, on 2016 August $21$$-$$22$. We used a 1200\,line\,mm$^{-1}$ grating with a 0.86\arcsec\ slit, yielding a spectral resolution of 1.3\,\AA. The optimally extracted \citep{1986PASP...98..609H} spectra were wavelength calibrated using sky emission lines and rebinned to a heliocentric frame using the {\sc pamela} and {\sc molly} packages \citep{1989PASP..101.1032M}; the signal-to-noise (S/N) per resolution element in Table~\ref{tab:rv206197016} is calculated at 6400\,\AA. Using the period and ephemeris defined in Table~\ref{tab:badflux}, our observations covered Phases\,$0.82-0.92$ and $0.20-0.27$, respectively. We fit a two-component Gaussian to find the radial velocity for each averaged spectrum, and see marginal evidence for shifts; however, our data do not definitely confirm velocity changes caused by a close companion to EPIC\,206197016.

% +++++++++++++++++++++++++++++ Table 2 +++++++++++++++++++++++++++++ %
\begin{table}
 \centering
  \caption{Radial velocity measurements of H$\alpha$ using SOAR/Goodman of the possible 19.898\,hr binary, EPIC\,206197016 \label{tab:rv206197016}}
  \begin{tabular}{@{}lcccr@{}}
  \hline
Time (BJD$_{\rm TDB}$)  & Airmass  & Exposures     & S/N & RV (km s$^{-1}$)   \\
\hline
2457621.68101    &  1.13  & 7$\times$480\,s & 26 & $+$47(24)       \\
2457621.71928    &  1.07  & 7$\times$480\,s & 29 & $+$41(20)       \\
2457621.76120    &  1.08  & 8$\times$480\,s & 27 & $+$36(27)       \\
2457622.82748    &  1.26  & 4$\times$420\,s & 20 & $+$1(32)       \\
2457622.88382    &  1.72  & 7$\times$480\,s & 28 & $-$20(20)       \\
\hline
\end{tabular}
\end{table}
% ==================================================================== %

\subsection{Magnetism}
\label{sec:magnetism}

Previous studies have found that strongly magnetic ($>$$1$\,MG) white dwarfs show large-amplitude photometric variability on timescales of hours to days (e.g., \citealt{2004MNRAS.348L..33B,2013ApJ...773...47B}), in line with the distribution of asteroseismically derived white dwarf rotation periods \citep{2015ASPC..493...65K}. Most of these objects have effective temperatures $<$$10\,000$\,K, where their atmospheres should be convective, with variations typically attributed to spots.

All four of the photometrically selected white dwarf candidates shown in Figure~\ref{fig:badfluxnospec} with large-amplitude flux variations in {\em K2} have photometric colours consistent with effective temperatures $<$$9000$\,K, suggesting they should have fully convective atmospheres. We estimate the effective temperature for each in Table~\ref{tab:badflux} by comparing the ($u$$-$$g$,$g$$-$$r$) colours to Figure~1 of \citet{2014ApJ...796..128G}. The folded light curves of these apparently spot-modulated white dwarfs, shown in Figure~\ref{fig:badfluxnospec}, correspond to rotation periods of $1.23-8.31$\,d. Notably, EPIC\,210609259 (in the top right of Figure~\ref{fig:badfluxnospec}) has a light curve that can be well approximated by a white dwarf with a magnetic dipole with polar spots, with a rotation/observer inclination of $45\pm8$\degr\ and a rotation/magnetism colatitude of $38\pm10$\degr, linearly offset by $a_z = -0.31\pm0.04$.

In addition to these likely cool, convective white dwarfs with apparent spots, we see multiple hotter, strongly magnetic white dwarfs with large-amplitude variability. All have \teff\ $>$$30\,000$\,K, so their atmospheres should be radiative. However, Zeeman features can change in depth and shape as a function of rotation phase and induce variability, as seen in the strongly magnetic, $>$$45\,000$\,K white dwarf RE\,J0317$-$853 \citep{1999ApJ...510L..37B}.

EPIC\,211995459 (SDSSJ084330.81$+$201049.1) is a $60\,000$\,K magnetic DA white dwarf \citep{2016MNRAS.455.3413K}. It appears to have a similar spot geometry to EPIC\,210609259 (and similar $2$-d rotation period), but features a bright spot rather than a dark one. The shape of the modulation is also very similar to the bright spot on the hottest pulsating DB known, PG\,0112$+$104 \citep{2017ApJ...835..277H}.

Additionally, two white dwarfs observed with {\em Kepler} have complex spot modulation and rotation periods of roughly 6\,hr. The first, KIC\,9535405 (BOKS\,53856) was discovered in the original {\em Kepler} mission field; it is a DA with \teff\ $=34\,000$\,K with marginal evidence of a $\sim$350 kG magnetic field \citep{2011AJ....142...62H}. The other, EPIC\,211719918 (PG\,0853$+$164), has a similar effective temperature, $34\,520$\,K \citep{2013ApJS..204....5K}, and is a known weakly magnetic, variable DBA white dwarf \citep{1997ASSL..214..413P}. Previous studies have put the effective temperature of this white dwarf near the DBV instability strip, where it may pulsate from a helium partial-ionization zone \citep{2001ApJ...554.1118W}. Using 58.8\,s short-cadence {\em K2} data, we are able to improve limits on the lack of pulsations by an order of magnitude, ruling out any variability from $120-2000$\,s with semi-amplitudes above 0.12\,ppt in PG\,0853$+$164.

\subsection{Pulsations}
\label{sec:pulsations}

Non-radial oscillations have been observed for more than half a century in white dwarfs, which cause optical variations with amplitudes exceeding 1\,per\,cent at periods from $100-1400$\,s \citep{2008PASP..120.1043F}. Pulsating white dwarfs are bad flux standards. We have removed all pulsating white dwarfs from our sample; they will be discussed in detail in future manuscripts.

However, a new outburst phenomenon occurring at the cool edge of the DAV instability strip deserves special mention. These brightening events, which recur stochastically on day-to-week timescales, can brighten a white dwarf by more than 40\,per\,cent for several hours \citep{2015ApJ...810L...5H}. The first six outbursting white dwarfs all have flux excursions in excess of 10\,per\,cent, each event lasting several hours \citep{2016arXiv160909097B}. So far, we have only observed this phenomenon in the coolest DAVs \citep{2016ApJ...829...82B}.

Outbursts may be the result of a transfer of pulsation energy into heating the star, possibly from nonlinear mode coupling \citep{2015ApJ...810L...5H}. This suggests the phenomenon likely happens among the other white dwarf instability strips. Data from the original {\em Kepler} mission may bear this out: the central star of the planetary nebula Kr61 (KIC\,3231337) was observed to show stochastic, several per\,cent brightening events every few days \citep{2015MNRAS.448.3587D}. Analysis of short-cadence {\em Kepler} photometry show this is indeed a pulsating white dwarf with relatively long ($>$750\,s) oscillation periods at the cool edge of the DOV instability strip. Outbursting white dwarfs make for especially bad flux standards.

\section{Discussion and Conclusions}
\label{sec:discussion}

We have empirically assessed the viability of white dwarfs as flux standards by analyzing the stability of nearly 400 non-pulsating, apparently isolated white dwarfs observed by the {\em Kepler} spacecraft through {\em K2} Campaign 8. Our results confirm that the vast majority ($>$$97$\,per\,cent) of white dwarfs are suitable flux standards; key caveats to rule out are pulsations, binarity, and magnetism. Only nine white dwarfs in this sample show coherent photometric variability on 0.04$-$10\,d timescales with amplitudes exceeding 1\,per\,cent, detailed in Table~\ref{tab:badflux}. Additional groups have set out to analyze white dwarf stability at even-lower, mmag levels using {\em K2} photometry of brighter targets (Z. Xue \& B. Schaefer, private communication).

Observers can avoid pulsating white dwarfs by not using those with effective temperatures near the empirical DAV and DBV instability strips, which correspond to the onset of convection for hydrogen- and helium-atmosphere white dwarfs, respectively. This occurs between roughly $12\,500-10\,500$\,K for canonical-mass DAVs \citep{2015ApJ...809..148T} and roughly $32\,000-20\,000$\,K for canonical-mass DBVs \citep{2009ApJ...690..560N}. The DOV instability strip occurs for white dwarfs $>$$100\,000$\,K; we recommend against such hot objects for reasons of binarity.

Observers can avoid most binary white dwarfs by searching for line-of-sight companions, commonly M dwarfs \citep{2016MNRAS.458.3808R}. However, our {\em K2} results suggest that the hottest white dwarfs (near $\sim$$100\,000$\,K) can easily outshine low-mass companions. Since it is difficult to detect close companions, it is thus difficult to assess whether such a hot star is a reliable flux standard. DO white dwarfs are also bad flux standards: more than 10\,per\,cent of planetary nebulae nuclei show photometric variations from a close companion \citep{2000ASPC..199..115B,2015AJ....150...30H}.

We find that spot modulation from magnetic white dwarfs is the most difficult caveat to rule out when seeking a reliable flux standard. The high surface gravity of a white dwarf significantly broadens any absorption lines present, so Zeeman splitting is typically undetectable for global fields below $\sim$$1$\,MG without high-resolution spectroscopy \citep{2013MNRAS.429.2934K}. 

Recently, spots have been detected in multiple white dwarfs with relatively firm upper limits on surface magnetic fields. \citet{2015ApJ...814L..31K} discovered a massive white dwarf with 38-min flux modulation exceeding 6\,per\,cent amplitude, but put an upper limit on the magnetic field of $<$$70$\,kG. More stringently, \citet{2017ApJ...835..277H} discovered a bright spot on the hot DBV PG\,0112$+$104 exceeding $>$$0.25$\,per\,cent amplitude, but symmetry in the observed pulsations require a global field $<$$10$\,kG. Empirically, variability from spot modulation is not reserved for purely convective white dwarfs, nor for strongly magnetic white dwarfs.

Although we show that the chances are low that a non-pulsating, isolated white dwarf has high-amplitude, intrinsic variability, we also show it is difficult to pre-screen against spot modulation from photometry or spectroscopy. Our results suggest the need to empirically assess the stability of a white dwarf before relying on it as an absolute flux standard, especially the anchors for flagship-class space missions such as {\em JWST}. Such empirical efforts are underway for {\em Gaia} calibration (e.g., \citealt{2016MNRAS.462.3616M}).

For example, future multi-epoch light curves from the high-precision photometry produced by {\em Gaia} will allow an empirical determination of the flux stability of hundreds of thousands of white dwarfs. These objects, as well as those shown empirically to be constant from {\em Kepler} observations, should form the basis of future networks of flux standards. We will publish our full catalog of constant white dwarfs at the end of the {\em K2} mission, which could continue beyond Campaign 17.

White dwarfs are intrinsically stable enough to highlight long-timescale instrumental artifacts from {\em Kepler}, especially the rolling bands that affect many of the CCDs on the spacecraft. Figure~\ref{fig:rollingbands} shows the light curves of two faint targets affected by this electronics noise, and highlights the need to rule out instrumental artifacts when analyzing the faintest targets observed in {\em K2} for intrinsic astrophysical variability.

\section*{Acknowledgments}

We wish to acknowledge the many {\em K2} Guest Observer proposers for ensuring these white dwarfs be observed from space, including teams led by R. Alonso, M. R. Burleigh, Steven D. Kawaler, M. Kilic, and Avi Shporer. Support for this work was provided by NASA through Hubble Fellowship grant \#HST-HF2-51357.001-A, awarded by the Space Telescope Science Institute, which is operated by the Association of Universities for Research in Astronomy, Incorporated, under NASA contract NAS5-26555. The research leading to these results has received funding from the European Research Council under the European Union's Seventh Framework Programme (FP/2007-2013) / ERC Grant Agreement n. 320964 (WDTracer). Based on observations obtained at the Southern Astrophysical Research (SOAR) telescope, which is a joint project of the Minist\'{e}rio da Ci\^{e}ncia, Tecnologia, e Inova\c{c}\~{a}o da Rep\'{u}blica Federativa do Brasil, the U.S. National Optical Astronomy Observatory, the University of North Carolina at Chapel Hill, and Michigan State University.

{\it Facilities:} {\em Kepler}, {\em K2}, SOAR, SDSS

\bibliographystyle{mnras}
\bibliography{hermes_wdflux}

\label{lastpage}
\end{document}